# Ranking protein-protein models with large language models and graph neural networks

Xiaotong Xu[1], Alexandre M.J. J. Bonvin[1], *

[1] Computational Structural Biology Group, Department of Chemistry, Bijvoet Centre for Biomolecular Research, Faculty of Science, Utrecht

**Abstract** Protein–protein interactions (PPIs) are associated with various diseases, including cancer, infections, and neurodegenerative disorders. Obtaining three-dimensional structural information on these PPIs serves as a foundation to interfere with those or to guide drug design. Various strategies can be followed to model those complexes, all typically resulting in a large number of models. A challenging step in this process is the identification of good models (near-native PPI conformations) from the large pool of generated models. To address this challenge, we previously developed DeepRank-GNN-esm, a graph-based deep learning algorithm for ranking modelled PPI structures harnessing the power of protein language models. Here, we detail the use of our software with examples. DeepRank-GNN-esm is freely available at https://github.com/haddocking/DeepRank-GNN-esm.

**Keywords** Protein-protein scoring, graph neural network, protein language model, DeepRank-GNN-esm

## 1 Introduction

Protein–protein interactions (PPIs) are fundamental for driving various biological processes within cells, such as cell growth, metabolism, and regulatory pathways. Obtaining three-dimensional (3D) structural insights into these PPIs is vital for understanding both normal physiological processes and their disruption in case of disease, guiding efforts toward the development of drugs to restore a normal functioning. Next to experimental structural methods such as cryo-electron microscopy and X-ray crystallography, various computational techniques have emerged as alternatives, ranging from integrative modelling [1–3] to artificial intelligence (AI)-based methods [4–6].

Despite recent progress in predicting the structure of biomolecular complexes, accurately identifying near-native conformations (good models) from the large pool of generated models, which is known as the 'scoring problem,' remains challenging as demonstrated for example in the Critical Assessment of Predicted Interactions (CAPRI) [7, 8]. To address this issue, we have previously developed DeepRank-GNN [9], a deep learning architecture that enables analysis of PPI interfaces using graph neural networks. The workflow of DeepRank-GNN is as follows: Given an input PPI 3D model, it converts the interaction interface into a residue graph where nodes represent protein residues at the interface and edges denote interactions between them. All graphs are stored and handled using the HDF5 format [10] to ensure smooth I/O operations and efficient training. The generated graph is then processed through a pre-trained graph neural network and the model quality is predicted in terms of $f_{nat}$, a metric between 0 and 1 reporting the fraction of native intermolecular contacts with respects to the ground truth (which would be the experimental structure of the complex if available). The network has been trained on large datasets alongside their corresponding $f_{nat}$ values, learning to rationalize between 3D structure, sequences, and interface quality. In the original DeepRank-GNN approach, five node features are used to capture various aspects of protein-protein interaction interfaces. These include residue type, polarity, buried surface area, and charge, which aims to characterize the physical properties of the interface. Additionally, we integrate evolutionary information through Position-Specific Scoring Matrix (PSSM) profiles. PSSM features have been shown to play a crucial role in improving the predictive performance of DeepRank-GNN [9]. They are however quite computational demanding to generate in terms of time and resources, requiring typically large sequence alignments to extract the evolutionary information.

The field of natural language processing has witnessed remarkable advancements, largely driven by the escalating scale of large language models (LLMs) [11]. This progression has extended to the realm of protein science with the development of sophisticated protein language models such as ProteinBERT [12], ProGen [13], Evolutionary Scale Modeling-2/3 (ESM-2/EMS-3) [6, 14]. ESM-2 has been trained to predict the identity of randomly masked amino acids in a protein sequence. By solving billions of sequence puzzles, ESM-2 effectively captured evolutionary insights spanning across sequences. Consequently, sequence embeddings extracted from ESM-2 models could provide valuable features for down-stream PPI-related tasks such as protein structure prediction [15], functional annotation [16] or generating novel sequences for protein families [17]. Recognizing the great potential of such large language models, we have replaced the PSSM features with embeddings from the ESM-2 protein language model, resulting in a new version of our software called DeepRank-GNN-esm [18]. This improvement greatly reduced the computation time at no cost in-, or even better performance than the original PSSM-based model on two applications: 1) Discriminating physiological from non-physiological interfaces and 2) Ranking docked PPI models.

In this chapter, we provide comprehensive instructions for the use of DeepRank-GNN-esm, focusing on its application for scoring PPI conformations. We introduce two execution modes for the algorithm with clear guidance on their implementations. Additionally, we dive into the process of customizing the deep learning architecture to meet individual requirements, ensuring flexibility and adaptability to diverse research needs.

## 2 Materials

The following resources are required to use our deep learning tools.

### 2.1 Software requirements

1. ESMFold [6] software package (available at https://github.com/facebookresearch/esm): code and pre-trained weights for protein language model embedding calculation.
2. DeepRank-GNN-esm [18] software package (https://github.com/haddocking/DeepRank-GNN-esm ): code and pre-trained model weights for $f_{nat}$ computation.
3. Python package PDB-tools (version 2.5.0 https://github.com/haddocking/pdb-tools ): a set of useful tools for pre-processing input coordinate files from the Protein Data Bank (PDB) [19].
4. Conda (https://conda.io/projects/conda/en/latest/index.html): An open-source software environment manager.
5. Jupyter Notebook (https://jupyter.org/) (optional): A web-based interactive computing platform.

### 2.2 Hardware requirements

We have tested the installation on CentOS 7 (with CUDA 11.6 and GeForce GTX 1080 Ti card), RockyLinux (with CUDA 12.5 and GeForce GTX 1080 Ti card), Ubuntu 20.04 LTS (with CUDA 12.4 and A100 GPU card) and openSUSE tumbleweed (with CUDA 12.3 and A4500 GPU card). A Linux computer with multiple CPUs and/or GPU is preferable.

## 2.3 Data requirements

Models of protein-protein complexes in PDB format.

## 3 Methods

In this section, we outline the process for evaluating the quality of PPI conformations using the DeepRank-GNN-esm algorithm. Additionally, we offer instructions on how our deep learning architecture can be retrained and/or re-designed for new applications.

To rank PPI conformations with DeepRank-GNN-esm, the following steps should be followed:

1. Install the required software packages.
2. Prepare and preprocess the input data (the models to be scored).
3. Run $f_{nat}$ prediction using the command line interface tool from a terminal or the DeepRank-GNN-esm package at python level.
4. Analyze results.

In the following, all those steps are described in details and illustrated with two examples using different settings, highlighting the versatility and efficacy of each approach.

## 3.1 Installation of the required software packages

DeepRank-GNN-esm can be used either as a command-line tool or a python package. Installation instructions for both settings are provided below:

1. Download and install Anaconda (https://docs.anaconda.com/free/anaconda/install/index.html)
2. Open a Linux terminal and clone the DeepRank-GNN-esm Github repository
   ```
   > git clone https://github.com/haddocking/DeepRank-GNN-esm
   > cd DeepRank-GNN-esm
   ```
3. Install the CPU or GPU version of DeepRank-GNN-esm depending on the hardware available in your system via conda:
   CPU version: Create and activate the conda environment
   ```
   > conda env create -f environment-cpu.yml
   ```

```
> conda activate deeprank-gnn-esm-cpu
```

GPU version: Create and activate the conda environment
```
> conda env create -f environment-gpu.yml
> conda activate deeprank-gnn-esm-gpu
```
4. Install the command line tool
   ```
   > pip install .
   ```
5. In the same installation directory, clone the ESMFold github repository (*see* **Note 1**)
   ```
   > git clone https://github.com/facebookresearch/esm.git
   ```
6. Install pdb-tools
   ```
   > pip install pdb-tools
   ```
7. Install Jupyer Notebook (Optional)
   ```
   > pip install notebook
   ```

Note that once you have installed all software, before using it in a new terminal session you will have to activate the conda environment again with:
```
> conda activate deeprank-gnn-esm-gpu
```

### 3.2 Preprocessing of the input data

Deeprank-GNN-esm is primarily tailored for interactions involving two chains. To run computation for complexes with more than two chains, refer to Note 2. Prior to utilizing the tool, renumbering the residue indexes for every chain in the PDB file is required (*see* **Note 2**). For this purpose, we recommend to use the provided script 'pdb_renumber.py': For example, for one input PDB file (e.g. input_protein.pdb), the script will renumber all the chains in the protein starting from residue '1' and generate a new PDB file in the user-defined output directory (note that if the output directory already exists it will be overwritten). To run it type:
```
> python scripts/pdb_renumber.py input_protein.pdb 1 output_dir
```

### 3.3 Run DeepRank-GNN-esm computation

#### 3.3.1 Run computation with command line interface tool

We use one example input to demonstrate the usage of the tool.

1. First create a separate 'tutorial' directory where we will run this example:
   ```
   > mkdir tutorial
   > cd tutorial
   ```
2. Download an example PDB file from the PDB (*see* Note 3)
   ```
   > pdb_fetch 1B6C > 1B6C.pdb
   ```
3. Prepare the input structure. This python command will overwrite the original PDB file.
   ```
   > python ../scripts/pdb_renumber.py 1B6C.pdb 1 .
   ```
4. Run $f_{nat}$ computation:
   To assess the quality of the protein-protein interface between chain 'A' and chain 'B', run the following command (*see* **Note 4**). This command requires four inputs from the user: the PDB file of interest ("`1B6C.pdb`"), the identifiers of the two chains you wish to assess (chain_id1 "`A`" and chain_id2 "`B`"), and the path to the pre-trained model weights (`$MODEL`).
   ```
   > export MODEL=../paper_pretrained_models/scoring_of_docking_models/gnn_esm/treg_yfnat_b64_e20_lr0.001_fold-all_esm.pth.tar
   > deeprank-gnn-esm-predict 1B6C.pdb A B $MODEL
   ```
   For a description of the resulting output refer to Section 3.4.

When ranking a small number of PDB structures (less than 100), we recommend using the command-line interface tool. This tool consolidates all functionalities into one user-friendly command. However, the process may be relatively slow because it handles only one input PDB at a time and running the predictions directly from Python (see next section) might be more efficient.

### 3.3.2 *Run the* computations in Python using the deep learning architecture

DeepRank-GNN-esm can also be used as a python deep learning architecture. We will use as an example (PDBID:1ATN) to demonstrate the process. By running parallel predictions harvesting multiple CPUs for data loading, we can accelerate the process compared to the command-line tool version.

Below are the steps for using the Deeprank-GNN-esm architecture for making $f_{nat}$ predictions with example PDB files (*See* **Note 5** for naming requirement for the PDB files) as input. For the convenience for the reader, we also provide a Jupyter notebook version of all the steps (*See* **Note 6)**.

1. From within the directory where you installed the software, create a new directory for the tutorial and copy all example PDB files to the new directory:
   ```
   > cd ../
   > mkdir tutorial_python
   > cp -r example/data/pdb/1ATN tutorial_python
   > cd tutorial_python
   ```

472. Prepare all input PDBs in the folder: Inside the `tutorial_python` directory, in bash shell, run:
   ```
   > for pdb_file in 1ATN/*.pdb; do python
   ../scripts/pdb_renumber.py "$pdb_file" 1 1ATN/; done
   ```

3. All input PDBs have been renumbered and organized in the folder "`1ATN`". To compute ESM-2 embeddings for them, we begin by extracting sequences from the PDBs (*see* **Note 7**). Open a terminal, activate the deeprank-gnn-esm conda environment (see above) and then type:

   ```
   > python ../scripts/get_fasta.py 1ATN/ A B
   ```

   This command generates a fasta file ("`1ATN.fasta`") from chain "`A`" and chain "`B`" for this complex. For other chain IDs, replace "`A`" and "`B`" with specific chain identifiers. To calculate the ESM-2 embeddings from the input fasta file, as described in the supplementary material in our original publication [18], we recommend using the script provided by ESM-Fold [6] (*see* **Note 8**):

   ```
   > mkdir embedding
   > python ../esm/scripts/extract.py
   esm2_t33_650M_UR50D 1ATN.fasta embedding --repr_layers
   33 --include mean per_tok
   ```

   The command generates ESM-2 embeddings for all sequences in "`1ATN.fasta`" with pre-trained ESM-2 model "`esm2_t33_650M_UR50D`" and stores them in the "`embedding`" directory.

4. To convert the input PDBs into interface graphs that can be processed and handled by the network, in the current directory, create a python file named '`prepare_HDF5.py`', copy the following python code snippet into the file:

   ```
   from deeprank_gnn.GraphGenMP import GraphHDF5
   pdb_path = "1ATN"
   embedding_path = "embedding"
   nproc = 20
   outfile = "1ATN_residue.hdf5"
   GraphHDF5(pdb_path = pdb_path,
             embedding_path = embedding_path,
               graph_type = "residue",
               outfile = outfile,
               nproc = nproc,
               tmpdir="./tmpdir")
   ```

   Run the python file in the '`tutorial_python`' directory with:
   ```
   > python ./prepare_HDF5.py
   ```

   This code will generate the input interface graphs with ESM-2 embeddings as one of their node features for all PDB files in the directory and compact them into the output HDF5 file format using 20 CPU cores.

5. To use pre-trained model to rank the input conformations in HDF5 format, create a new python file named ''`predict.py`' in the current directory and add the following code to the file:

```
from deeprank_gnn.ginet import GINet
from deeprank_gnn.NeuralNet import NeuralNet as NN
database_test = "1ATN_residue.hdf5"
gnn = GINet
target = "fnat"
edge_feature = ["dist"]
node_features=["type", "polarity", "bsa", "charge", "embedding"]
threshold = 0.3
pretrained_model="../paper_pretrained_models/scoring_of_docking_models/gnn_esm/treg_yfnat_b64_e20_lr0.001_foldall_esm.pth.tar"
device_name = "cuda:0"
num_workers = 10
model = NN(
            database_test,
            gnn,
            device_name = device_name,
            edge_feature = edge_feature,
            node_feature = node_features,
            target = target,
            num_workers = num_workers,
            pretrained_model=pretrained_model,
            threshold = threshold)
model.test(hdf5 = "GNN_esm_prediction.hdf5")
```

   Run the python file in the '`tutorial_python`' directory with:
```
> python ./predict.py
```

As outlined in the original publication, the available node features include `'type'`, `'polarity'`, `'charge'`, `'bsa'`, and `'embedding'`. It is important to note that since the model is trained with all these features, we highly recommend using all of them during inference time for optimal performance.

*3.4 Analysing the results*

**3.4.1 Analysing the output from the command line interface tool**

The command line tool generates a result folder within the current working directory. The structure of this folder is as follows:

```
1B6C-gnn_esm_pred_A_B
├── 1B6C.pdb
├── all.fasta
├── 1B6C.A.pt
├── 1B6C.B.pt
├── graph.hdf5
├── GNN_esm_prediction.hdf5
└── GNN_esm_prediction.csv
```

The output folder "1B6C-gnn_esm_pred_A_B" contains the pre-processed input protein ("1B6C.pdb"), the extracted protein sequences ("all.fasta"), the calculated ESM-2 embeddings for protein chain 'A' and chain 'B' ("1B6C.A.pt", "1B6C.B.pt"), the generated interface graph ("graph.hdf5") and the prediction output files ("GNN_esm_prediction.hdf5" and "GNN_esm_prediction.csv") with predicted $f_{nat}$ values for the interface. In our paper [18], we use a $f_{nat}$ cutoff of 0.3 to differentiate between good and bad models, with a predicted '$f_{nat}$' value higher than 0.3 indicating a good interface conformation and vice versa.

To access the predicted output, simply open the 'GNN_esm_prediction.csv' with a txt editor. To read node features and edge features of the generated interface graph ("graph.hdf5"), use to the following python code:

```python
import h5py
f = h5py.File("1B6C-gnn_esm_pred_A_B/graph.hdf5")

#view all node features
node_features = f["1B6C"]["node_data"].keys()
print(node_features)

#view node feature embeddings
embeddings = f["1B6C"]["node_data"]["embedding"][()]
print(embeddings)

#view edge feature distance
distance = f["1B6C"]["edge_data'"]['dis'][()]
print(distance)
```

### 3.4.2 Analysing the output from the Python Deeprank-GNN-esm network architecture

The provided code snippet above generates a output HDF5 file named 'GNN_esm_prediction.hdf5'. Similar to the outputs from the command line tool, you can access predicted $f_{nat}$ through python code:

```
import h5py
f = h5py.File("GNN_esm_prediction.hdf5","r+")
mol_names = f["epoch_0000"]["test"]["mol"][()]
fnats = f["epoch_0000"]["test"]["outputs"][()]
for mol_name, fnat in zip(mol_names, fnats):
  print(mol_name.decode(), fnat)
```

## *3.5 Customizing the deep learning network*

DeepRank-GNN-esm is designed to facilitate easy adaptation for your own applications. To re-train it for customized tasks, follow these steps:

1. Construct and format training and evaluation data sets
2. Define the training tasks. Two training mode are supported: regression ('reg') or classification ('class').
3. Define the learning parameters: Determine 'batch_size', learning rate 'lr' based on your system setup and hardware capacity (*see* **Note 9)**
4. Load the network
5. Retrain the network

We use the following code example to demonstrate the process:

```
from deeprank_gnn.NeuralNet import NeuralNet
from deeprank_gnn.ginet import GINet

database_train = 'train.hdf5'
database_eval = 'eval.hdf5'

task='reg'
batch_size=64
lr = 0.001
node_feature = ['type', 'polarity', 'bsa', 'charge', 'embedding']
  edge_feature = ['dist']
```

```
model = NeuralNet(database_train,
                  GINet,
                  node_feature=node_feature,
                  edge_feature=edge_feature,
                  target=target,
                    task=task,
                    lr=lr,
                    batch_size=batch_size,
                    database_eval = database_eval)

model.train(nepoch=50,
            validate=True,
            save_model='best',
            hdf5='output.hdf5')
```

Graphs, neural network layers and training target could also be customized. For advanced users, we refer you to the example here https://deeprank-gnn.readthedocs.io/en/latest/tutorial.advanced.html.

## 4 Notes

1. The environment required for ESMFold is integrated into the conda environment for DeepRank-GNN-esm. Users only need to clone the repository to access the helper scripts necessary for embeddings calculation.

2. To ensure correct matching between the calculated ESM-2 embeddings (shaped [AA_len, 1280]) and their corresponding residue nodes in the graph, we employ residue renumbering prior to computations. This approach will ensure the residue numbering in all chains is continuous and starts from residue '1'. As a result, the *ith* element in the embeddings corresponds to the *ith* residue in the sequence. It's worth noting that gaps in the structure do not affect this process, as long as the sequence order in the PDB file matches the sequence provided to ESM-2.

3. To use our scoring function with files in the default CIF format from the Protein Data Bank, you need to first convert them to the PDB format. We recommend using the '`pdb_fromcif`' command from PDB-tools for this conversion.

4. Deeprank-GNN-esm is primarily tailored for interactions involving two chains. While our method may not directly evaluate the overall quality of complexes

      with more than two chains, it can still be effectively applied to assess interactions between specific pairs of chains within such complexes. In such cases, providing the two relevant chain IDs will enable our method to function probably.

5. All input PDB file should be named according to the following format: PDBID (4-letter code) followed by any combination of characters of any length and ending with the '.pdb' extension.

6. A Jupyter notebook version of the tutorial (`tutorial.ipynb`) is distributed with the package. To run the tutorial, first install the conda environment and Jupyter Notebook. Then initiate Jupyter Notebook with command `jupyter notebook` and run the tutorial inside the DeepRank-GNN-esm installation folder.

7. We assume that all input PDB conformations share the same fasta sequence, which is typically the case in scoring scenarios in which the models are derived from one specific docking software. All modeled conformations for one complex should be stored in one directory. In such scenarios, it suffices to calculate the fasta sequence for only one PDB. However, if the input PDB conformations possess different sequences, it is necessary to generate fasta sequences for each of them.

8. Refer to the ESM-Fold repository or the Supplementary Material of our original DeepRank-GNN-esm publication for detailed explanations of each input parameter.

9. As an example of learning parameters, for a GPU GeForce GTX 1080 Ti card (11G memory), we use 'batch _size'=64 and 'lr'=0.001.

## Acknowledgements

This work was supported by the European Union Horizon 2020, project BioExcel [823830]. X.Xu acknowledges financial support from the China Scholarship Council, grant 202208310024.